\documentclass[eprint]{raa}
\usepackage{graphicx,times}
\usepackage{natbib}
\usepackage{amssymb,amsmath}
\bibpunct{(}{)}{;}{a}{}{,}

\usepackage[a4paper=true,pagebackref=true]{hyperref}
\hypersetup{pdftitle = The title of my PDF,
           pdfauthor = My name,
           pdfsubject= The subject,
           pdfkeywords = keyword1 keyword2 keyword3} 
\hypersetup{colorlinks = true,
             linkcolor = green,
             anchorcolor = red, 
             citecolor = blue, 
             filecolor = red, 
             pagecolor = red, 
             urlcolor = red}

\begin{document}

\title{ Boresight Alignment of DArk Matter Particle Explorer}

\author{ Wei Jiang\inst{1, 2}, Xiang Li\inst{1}\thanks{Correspondence to: \it xiangli@pmo.ac.cn},  Kai-Kai Duan\inst{1, 3}, Zhao-Qiang Shen\inst{1,3}, Zun-Lei Xu\inst{1}, Jing-Jing Zang\inst{1}, Shi-Jun Lei\inst{1} and Qiang Yuan\inst{1,2}}

\institute{ Key Laboratory of Dark Matter and Space Astronomy,
            Purple Mountain Observatory,
            Chinese Academy of Sciences,
            Nanjing 210008, China 
            \and
            School of Astronomy and Space Science, 
            University of Science and Technology of China, 
            Hefei 230026, China
            \and
            University of Chinese Academy of Sciences,
            Beijing 100049, China
}

\abstract{
The DArk Matter Particle Explorer (DAMPE) can measure $\gamma$-rays in the energy range from a few GeV to about 10 TeV. 
The direction of each $\gamma$-ray is reconstructed with respect to the reference system of the DAMPE payload. 
In this paper, we adopt a maximum likelihood method and use the $\gamma$-ray data centered around several bright point-like sources to measure and correct the angular deviation from the real celestial coordinate system, the so called ``boresight alignment'' of the DAMPE payload. 
As a check, we also estimate the boresight alignment for some sets of simulation data with artificial orientation and obtain consistent results. 
The time-dependent boresight alignment analysis does not show evidence for significant variation of the parameters.
}

\authorrunning{Wei Jiang, et al.}
\titlerunning{Boresight Alignment of DArk Matter Particle Explorer}
\maketitle

\section{Introduction}
\label{sect::intro}

The DArk Matter Particle Explorer (DAMPE) is a satellite mission which has been operating stably in a solar synchronous orbit since it was launched on December 17, 2015 from Jiuquan, China.
The payload is designed to detect high energy cosmic rays and $\gamma$-rays with high energy resolution and spatial resolution.
The detector consists of 4 sub-detectors \citep{CHANG2014,CHANG20176}, from the top to the bottom including a Plastic Scintillator Detector (PSD)  \citep{Yu2017,Ding2019}, a Silicon Tungsten tracKer-convertor (STK) \citep{Azzarello2016}, a BGO calorimeter (BGO) \citep{Zhang2016}, and a NeUtron Detectors (NUD) \citep{He2016}.
Each day, DAMPE collects about five million high energy particles, dominated by the protons, helium and heavier nuclei.
The on-orbit calibration with these data proves that DAMPE is fully operational as expected and all sub-detectors perform well.
One important part of the on-orbit calibration is the payload internal  and satellite alignment \citep{OnOrbit2018}.
The details of  alignments of the STK and PSD with the on-orbit data have already been published \citep{Tykhonova2018, Ma2018}. In this work we focus on the boresight alignment of DAMPE.

For each incident high energy particle, the direction is reconstructed with respect to the reference system of DAMPE payload.
To get the celestial coordinates of the detected high energy particles, the incident directions should be transformed to the celestial coordinate system from the payload coordinate system .

As usual, the transformation depends on both the GPS navigation system and star-trackers. 
Actually, there are two star tracks on the satellite. When the star-trackers calculate the attitude angle of the satellite,
the need to compare with the zenith and azimuth provided by the GPS. Only if the deviation between the results is less
than a special value, it is regarded as an effective measurement. The payload system is nominally designed the same as the satellite system.
As a result of thermal variations, acoustic vibrations, zero gravity fluctuations, and uncertainty in the orbital parameters and star-tracker pointing, small deviations from real pointing are expected to be introduced to the resulting  celestial coordinate.
In addition to causing a systematic shift between the observed position and the real one of a point-like source,
such a mismatch will also give rise to a distorted point spread function (PSF) profile.

 In order to effectively measure and correct such a bias, we take the GeV $\gamma$-ray data centered around a few brightest point-like sources detected by DAMPE to measure and then correct the angular deviation from the real celestial coordinate. Such a process is the  so called  ``boresight alignment'' of DAMPE payload.
In this work we introduce the method, result and verification of our boresight alignment in detail.

\section{Likelihood Analysis}
\label{sect::method}

The deviation needed to be calibrated can be described as a rotation in the payload coordinate system.
Therefore, with a rotation matrix the vectors in the payload coordinate and the satellite coordinate systems can be directly related. 
The rotation of a vector in the satellite coordinate system can be described by three Euler angles which are generally called roll$(\bm \psi)$,  pitch$(\bm \theta)$ and yaw$(\bm \phi)$ , which represent the rotation angles along x, y and z axes respectively.
The rotation matrix in terms of these variables reads
\begin{equation}
 \label{eq::tarnsmat}
\bm R(\bm \psi, \bm \theta, \bm \phi) \equiv
\begin{bmatrix}
1&0&0&\\
0&\cos \bm \psi & -\sin \bm \psi &\\
0&\sin \bm \psi & \cos \bm \psi &
\end{bmatrix}
\begin{bmatrix}
\cos \bm \theta&0&\sin \bm \theta &\\
0&1&0&\\
-\sin \bm \theta&0&\cos \bm \theta &
\end{bmatrix}
\begin{bmatrix}
\cos \bm \phi&-\sin \bm \phi&0&\\
\sin \bm \phi&\cos \bm \phi &0&\\
0&0&1&
\end{bmatrix}.
\end{equation}
Thus, the column vector  $\vec{\bm r}$ of the track in the payload coordinate system can be expressed as
\begin{equation}
\label{eq::track}
 \vec{\bm r} = \bm R(\bm \psi, \bm \theta, \bm \phi) \vec{\bm r}^\prime,
\end{equation}
where $\vec{\bm r}^\prime$ is the track constructed in the payload coordinate system before the boresight alignment.

The rotation transforms the position of the incident particles  in the payload coordinate, and subsequently modifies direction in celestial coordinate.
Based on the housekeeping data of navigation system and star trackers, the transformation at time $\bm t$ from the payload coordinate to the celestial coordinate can be computed as a matrix $\bm R_{\rm sky}(\bm t)$. The correction of a vector $\vec{\bm p}$ in celestial coordinates due to boresight alignment ($\bm \psi, \bm \theta, \bm \phi$) can be expressed as
\begin{equation}
\label{eq::orbit}
\vec{\bm p}(\bm t, \bm \psi, \bm \theta, \bm \phi) = \bm R_{\rm sky}(\bm t) \bm R(\bm \psi, \bm \theta, \bm \phi) \vec{\bm r}^\prime.
\end{equation}

Some bright GeV sources, such as Vela, Geminga and Crab, have been well measured in the radio and optical bands, and their positions in celestial coordinate system are precisely known.
Some bright GeV sources, such as Vela, Geminga and Crab, have been well measured in the radio and optical bands, and their positions in celestial coordinate system are
known with the precision of $\sim0.01arcsec$\citep{Fey2004}.

So, we can use the brightest $\gamma$-ray sources to carry out the boresight alignment.
A set of software packages (DmpST), which incorporate the instrument response functions (IRFs) and the science tools,  have been developed to analyze the $\gamma$-rays detected by DAMPE \citep{Duan2019}. In this work DmpST is adopted to achieve our purpose.
 Due to the point spread function (PSF) which is the probability distribution of the reconstructed direction for a $\gamma-ray$ event with specify primary energy 
and incident direction especially large at low energies and large incident angles (e.g., 68.3\% of the counts will be within $\sim0.8^\circ$ at 1 GeV and $50^\circ$) , \cite{Duan2019} 
   to analyze a single source the counts within a region around the source have to be included. We call that region the ``region of interest(ROI)''.

In order to determine the boresight alignment parameters with the on-orbit data, a likelihood maximization analysis of some bright $\gamma$-ray point-like sources has been preformed.
For a given source in a ROI, we have used a point-like $\gamma$-ray source model with a background component. 
To evaluate the background we consider all the other point-like sources in the ROI, Galactic diffuse emission, extra-galactic emission and some residual cosmic-ray background  in the ROI, which, for the sake of simplicity, is modeled as a uniform template with a power-law spectrum\citep{FerimiOnOrbit2009, Roth2012}.
After this model is multiplied by the exposure and convolved with the PSF of DAMPE, we get the expectation of the Poisson distributed variable observed by the $\gamma$-rays in the ROI.
We ignore the effect of misalignment in the exposure since the angular deviation are expected to be very small and the exposure changes smoothly in the ROI.
In the DmpST, we model the PSF by two King functions \citep{Duan2019} and it is consistent with the GEANT4 simulation for DAMPE, so we use it to describe the angular distribution of the selected $\gamma$-rays sample.
Thus, the unbinned Poisson likelihood \citep{Cash1979} of the reconstructed $\gamma$-ray directions to estimate the best fit parameters of the boresight alignment($\bm \psi, \bm \theta, \bm \phi$) is given as
\begin{equation}
\label{eq::likelihood}
{\rm ln}\,\mathcal{L}( \lambda, \bm{ \psi,  \theta,  \phi}) =
- \iiint_{\rm ROI} \sum_{\rm j=1}^{n_{\rm sources}} r_{j}\left (E', \vec{\bm p}; t', \vec{\lambda}_{j}
 \right ) + \sum_{\rm i=1}^{n_{\rm events}} {\rm ln} \left (
 \sum_{\rm j=1}^{n_{\rm sources}} r_{j} \left (
 E_{i}', \vec{\bm p}_{i}' ( {\bm \psi,  \theta,  \phi}); t', \vec{\lambda}_{j}
 \right ) \right ),
\end{equation}
where $\lambda(E, \vec{\bm p})$ is the expected contribution from the point-like source and background to the $\gamma$-rays with energy $E$ and direction $\vec{\bm p}$ in celestial coordinate system, $\vec{\bm p}'_{\rm i}$ is the direction of i-th photon after boresight alignment and $t_{\rm i}$ is the time when i-th photon is recorded by the detector. 
The {\tt MINUIT}\footnote{https://github.com/iminuit/iminuit} algorithm is adopted to perform the optimization \citep{James1975} and the three boresight alignment parameters are derived.

\section{Validation}
\label{sect::simu}

Besides the DmpIRFs and the analysis tools, DmpST contains a simulation module which is used to simulate photons from astrophysical sources and process those photons according to the DmpIRFs  during the on-orbit operation of DAMPE.

This simulation is cross-validated by the analysis tools in the DmpST \citep{Duan2019} which are used to fit the parameters of the classified $\gamma$-ray sources.
To validate the alignment method, we manually import a misalignment defined by the  Eq.~\ref{eq::orbit} to simulate a sample of $\gamma$-ray events,  then the same alignment approach is applied to the misaligned  sample.

In this test, we randomly select 6 groups of boresight parameters in the range between $-2^{\circ}$ to $2^{\circ}$, as summarized in the Table ~\ref{tab::simupara}.

\begin{table}[h]
\centering
\begin{tabular}{|c|c|c|c|c|c|c|}
\hline
 & \multicolumn{3}{|c|}{Preset Parameters} &  \multicolumn{3}{|c|}{Fitted Means and RMSs} \\\hline
Group &$ \bm \psi(^{\circ})$ & $\bm \theta(^{\circ})$ & $\bm \phi(^{\circ})$ &$ \bm \psi(^{\circ})$ & $\bm \theta(^{\circ})$ & $\bm \phi(^{\circ})$   \\\hline
1 & -0.3083 & -0.4682 & -1.7161 & 
-0.3068 $\pm$ 0.00246 & -0.4605 $\pm$ 0.00302 & -0.3046 $\pm$ 0.00434 \\\hline
2 & -0.7615 & -0.6123 &  0.3076 & 
-0.7613 $\pm$ 0.00236 & -0.6134 $\pm$ 0.00317 & -0.3078 $\pm$ 0.00401 \\\hline
3 & -1.1194 & -0.8551 & -0.0436 &
-1.1195 $\pm$ 0.00262 & -0.8562 $\pm$ 0.00275 & -0.0450  $\pm$ 0.00477 \\\hline
4 & -1.7765 &  1.2667 &  0.3106 & 
 -1.7743 $\pm$ 0.00275 & 1.2600 $\pm$ 0.00310 &  0.3084 $\pm$ 0.00492 \\\hline
5 &  0.8902 & -1.1783 & -0.5767 & 
0.8913 $\pm$ 0.00261 & -1.1771 $\pm$  0.00292 & -0.5746 $\pm$ 0.00406 \\\hline
6 &  0.1250 &  0.0211 & -0.1436 & 
0.1242 $\pm$ 0.00247 & 0.0222 $\pm$ 0.00313 & -0.1428 $\pm$ 0.00438 \\\hline
\end{tabular}
\caption{The preset boresight parameters and the statistic of the fitted reults for simulation samples}\label{tab::simupara}
\end{table}

\begin{figure}
\center
    \includegraphics[width=0.9\textwidth]{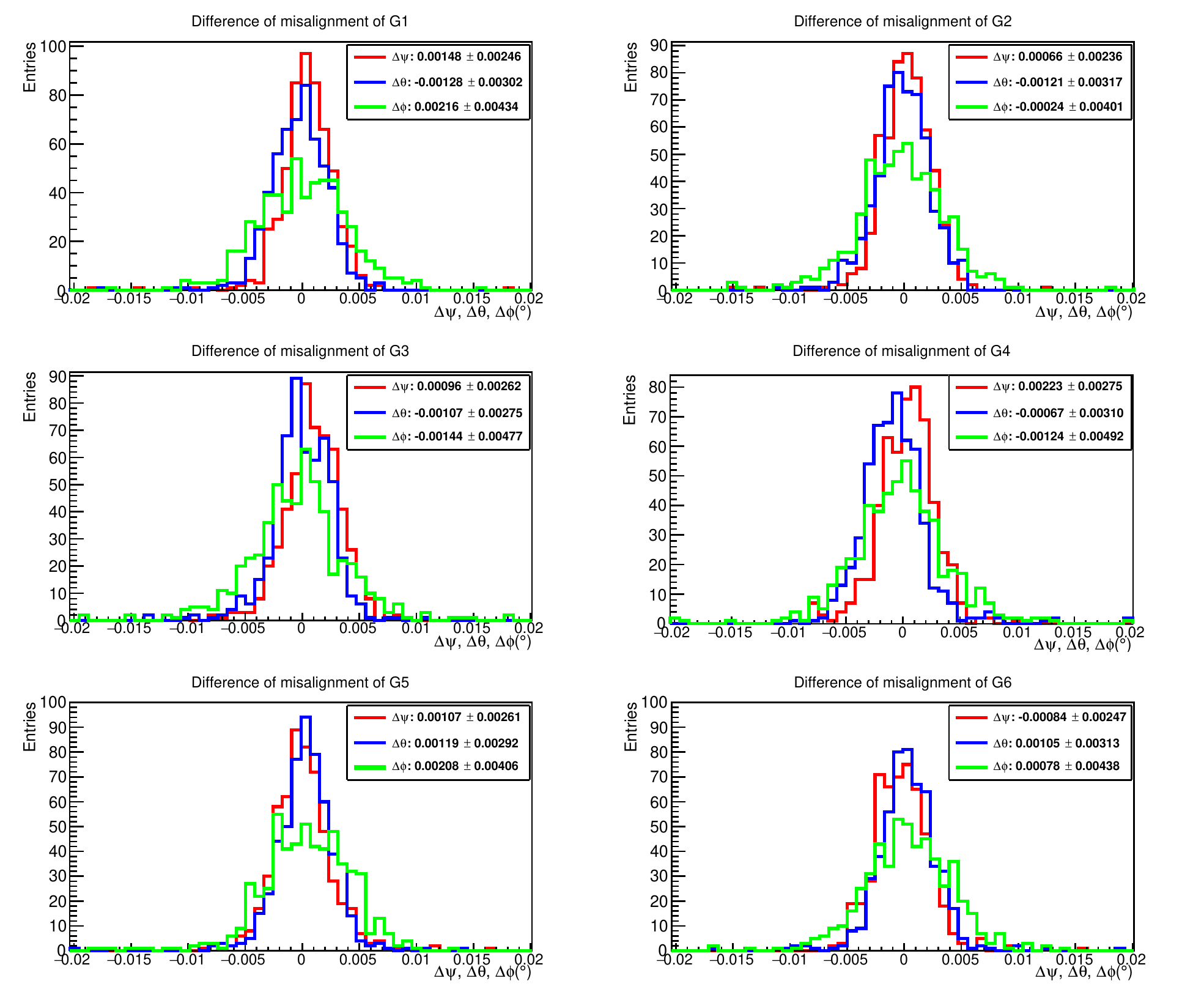}
\caption{The difference between the preset alignment parameters and the best-fitted results of the simulation data. The tiny amounts of deviation for all tests validate our boresight alignment method. }
\label{fig::simudiff}
\end{figure}

For each group of preset boresight alignment  parameters, we simulate 3 year observations to Vela pulsar  for 600 times.  We then analyze these $\gamma$-ray data with DmpST and then carry out the boresight alignment analysis with the procedure introduced in Section~\ref{sect::method}.
As shown in Tab. ~\ref{tab::simupara} and Fig.~\ref{fig::simudiff}, the difference between the preset alignment parameters and the best fitted ones are very small (the RMS is less than 0.005$^{\circ}$). 
Such results demonstrate that the boresight alignment method is reliable.

\section{Boresight Alignment of flight data}
\begin{figure}
\center
    \includegraphics[width=0.9\textwidth]{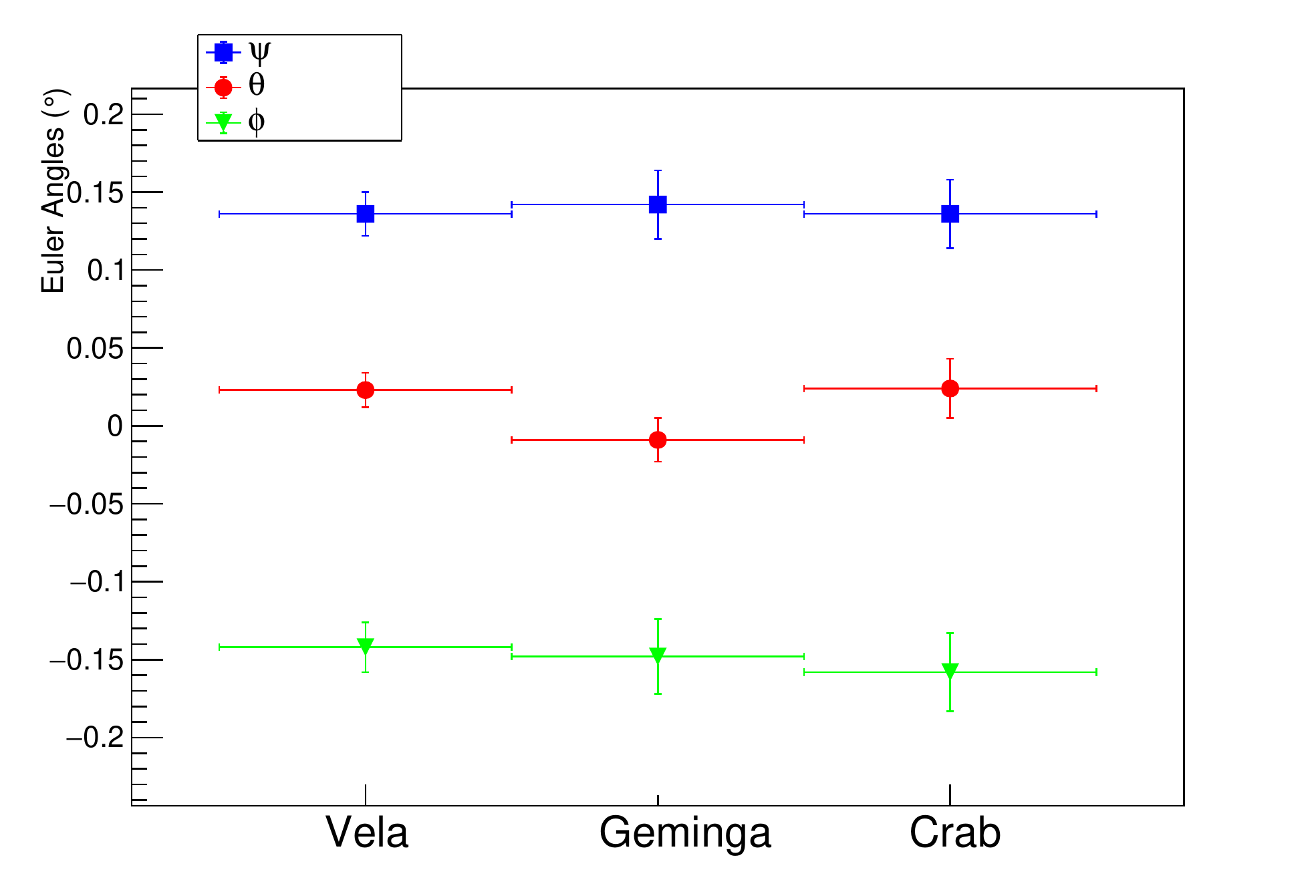}
\caption{The derived boresight alignment parameters for Vela, Geminga and Crab pulsars.}
\label{fig::compare}
\end{figure}

During the first 3 years of on-orbit operation, DAMPE successfully observed some bright GeV $\gamma$-rays   sources \citep{Xu2018, Liang2017}.
A few sources with accurate location information are bright enough for our purpose. The sources included in our following analysis are Vela, Crab and Geminga \citep{3FHL}.
The data detected from 2016 January 1 to 2019 January 1 are included in the analysis.
We select the photons within 4$^\circ$ from the targets and restrict the energy between 3 GeV and 100 GeV\citep{Duan2019}.
We fix the spectral parameters of the target sources to the values from the third Fermi-LAT catalog of high-energy sources \citep{3FHL} and optimize the rotation angles as well as the spectral parameters of the background using maximum likelihood method.
The resulting boresight alignment parameters for these three sources are consistent with each other (see Fig.~\ref{fig::compare}), as expected. In practice, the parameters which are applied for $\gamma$-ray analyses mainly come from the contribution of the Vela pulsar, which are 
\begin{equation}
\begin{aligned}
 \bm \psi = ~~0.136^\circ \pm 0.014^\circ ,\\
\bm \theta = ~~0.023^\circ \pm 0.012^\circ,\\
\bm \phi = -0.142^\circ \pm 0.018^\circ,
\end{aligned}
\end{equation}\label{eq::results}
Because the collected statistic of the Vela pulsar is several times larger than any other brightest $\gamma$-ray source, while the results of others can be validations and supplements. 
The uncertainties consist of two independent parts: the statistical errors fitted by Minuit and the measuring errors of $\sim0.01^\circ$ to each parameter given by

the engineering parameters of the satellite.

\begin{figure}
\center
    \includegraphics[width=1.0\textwidth]{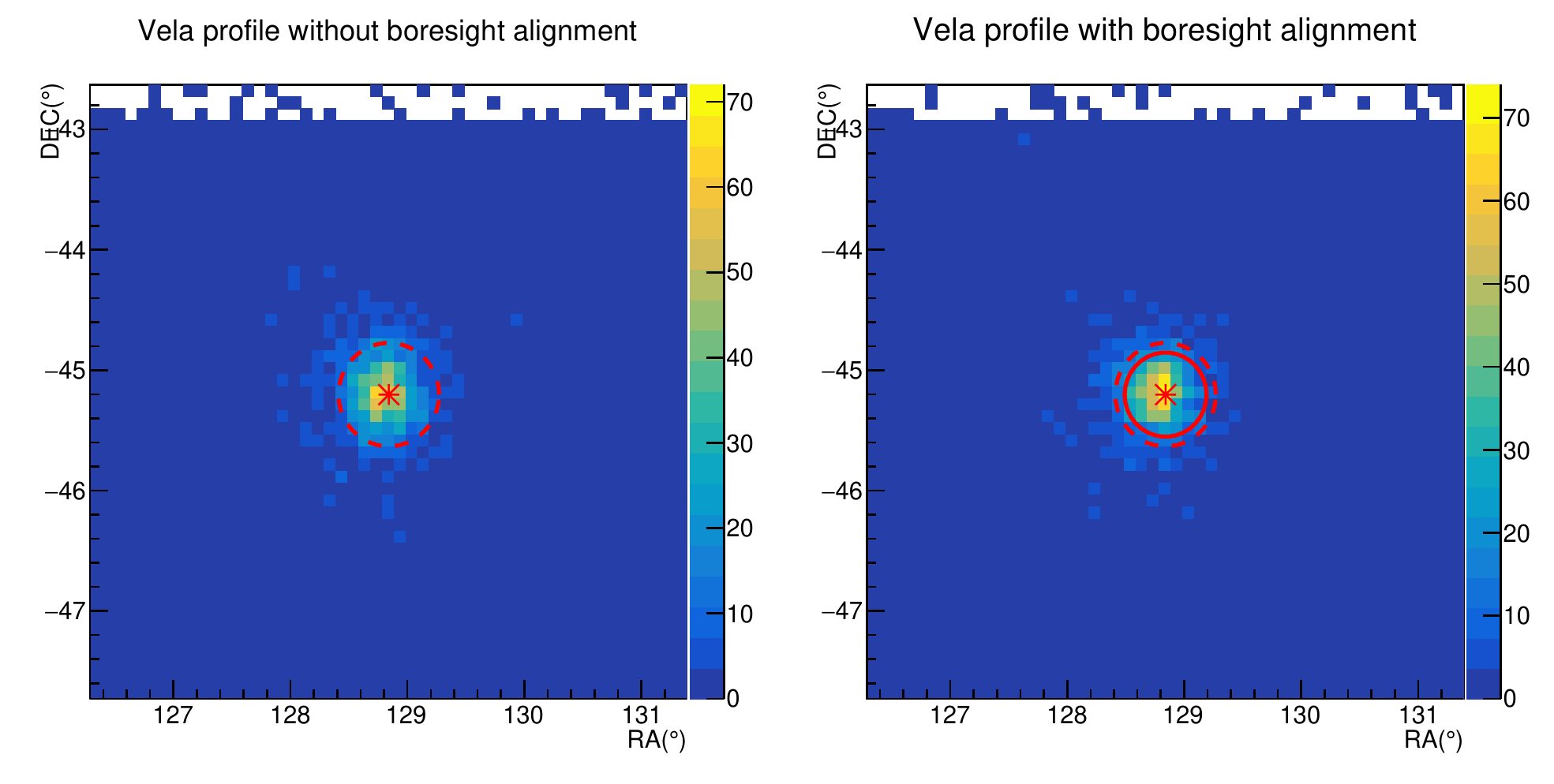}
\caption{The position profile of Vela pulsar with and without boresight alignment.
 The red contours encircle regions in which the 68.3\% of the events from the fitted point-like source.
    The dotted and solid line present the fitted result without and with boresight alignment respectively.
    The anglar radius of the 68.3\% contour decreases $\sim0.08^\circ$ after boresight alignment. }
\label{fig::profile}
\end{figure}

\begin{figure}
\center
    \includegraphics[width=0.9\textwidth]{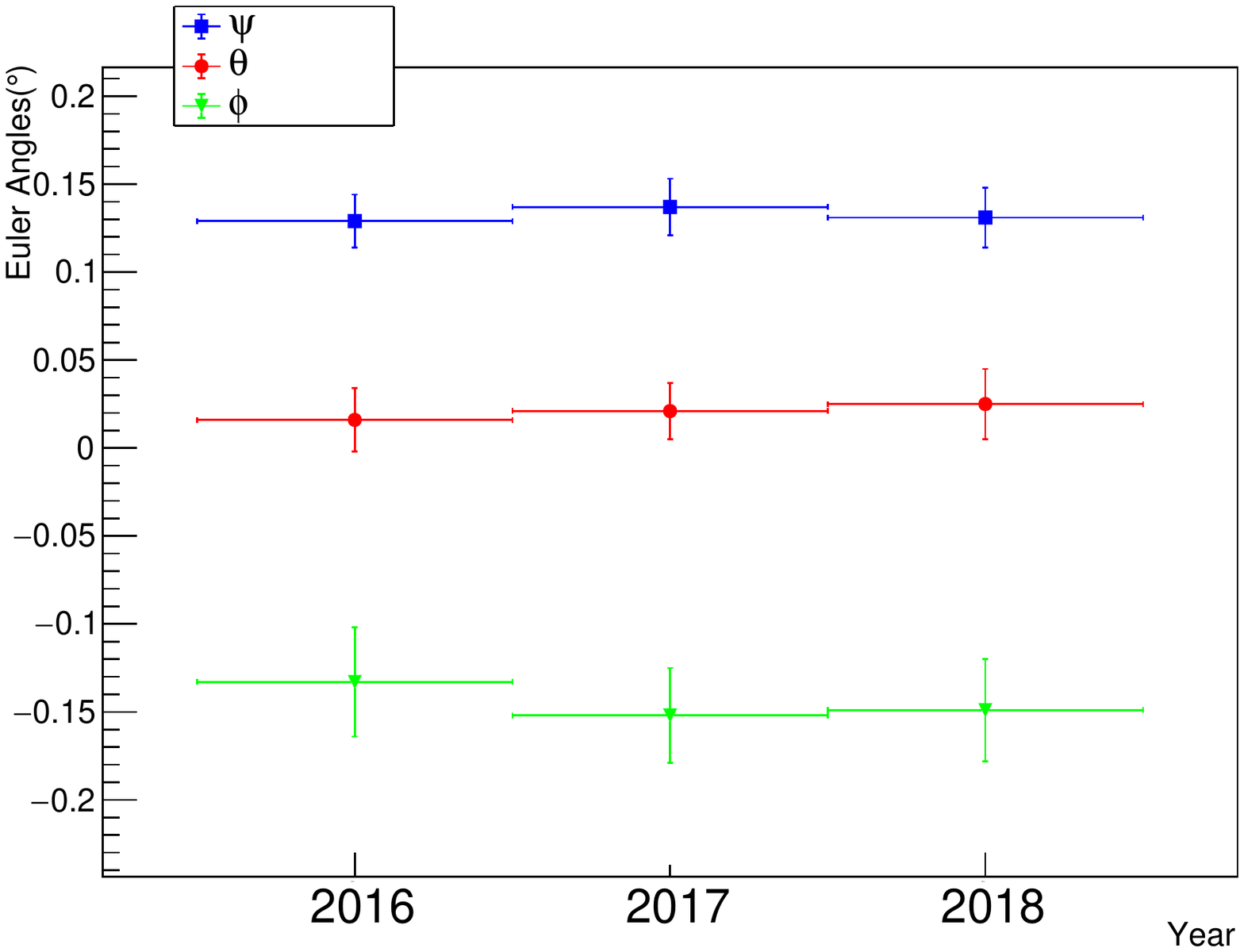}
\caption{The boresight alignment parameters of the Vela pulsar for each year. There is no evidence for the variation.  }
\label{fig::variation}
\end{figure}

After applying these boresight alignment, the position profile of the target source gets improved (see Fig.~\ref{fig::profile} for illustration). We have also studied the stability of the boresight alignment parameters by evaluating the parameters each year. As is shown in Fig.~\ref{fig::variation}, there are no significant variation in the boresight alignment parameters.

%

\section{Conclusion}
\label{Conclusion}
The payload of DAMPE consists of four sub-detectors and the directions of the incident particles are mainly reconstructed with the STK and BGO data. Such  directions are with respect to the reference system of the DAMPE payload  and further adjustments are needed to remove the miss match between the reconstructed direction and the intrinsic one.   For such a purpose we have developed a maximum likelihood method. Our  approach has been verified with the simulation data provided by DmpST.
We then take the 3 years of $\gamma$-ray data of Vela, Geminga and Crab pulsars measured by DAMPE to estimate the boresight alignment parameters. The fitted results of the Vela pulsar show that there is a primary offset between the orientation of payload and  satellite platform of $\sim0.15^\circ$, which is consistent with the other brightest $\gamma$-ray point-like sources.
We have also examined the variation of these parameters over the time and do not find any evidence for the evolution, which provides additional support to the stability of the sub-detectors of DAMPE in space, as found in the on-orbit calibration of the whole payload. With the boresight alignment corrections, the directions of the incident particles get improved.

\normalem
\begin{acknowledgements}
The DAMPE mission was founded by the strategic priority science and technology projects 
in space science of the Chinese Academy of Sciences (No. XDA04040000 and No. XDA04040400).
This work is supported in part by by National Key Program for Research and Development (No. 2016YFA0400200), 
the National Basic Research Program (No. 2013CB837000), the Strategic Priority Research Program 
of the Chinese Academy of Sciences ``Multi-Waveband Gravitational Wave Universe''(No. XDB23040000),
the Strategic Priority Research Program of theCAS(No. XDB23040000), 
Youth Innovation Promotion Association of CAS, 
the National Natural Science Foundation of China (Nos. U1738123, U1631111), 
the 100 Talents program of Chinese Academy of Sciences
and the Young Elite Scientists Sponsorship Program.
In Europe DAMPE activities receive generous support by the Swiss National Science Foundation (SNSF), Switzerland and the National Institute for Nuclear Physics (INFN), Italy.
\end{acknowledgements}

\bibliographystyle{raa}
\bibliography{ms}

\end{document}